\documentclass[12pt]{amsart}
\usepackage{graphicx}
\usepackage{geometry}
\usepackage{amssymb}
\usepackage{setspace}
\usepackage[dvipsnames]{pstricks}
\usepackage{pst-node}
\usepackage{pst-slpe}
\usepackage{pst-tree}
\usepackage{bbm}
\usepackage{amsmath, amsthm, amssymb}
\usepackage{pgfplots}
    \pgfplotsset{width=8cm,compat=1.5.1}
\usetikzlibrary{arrows.meta}
    \usetikzlibrary{shapes,snakes}
\usepackage{amsmath,amssymb,amsthm}
\usepackage{tikz}
\usepackage{caption}
\usepackage{subcaption}
\usepackage{booktabs}
\usepackage{empheq}
\usepackage[most]{tcolorbox}
\usepackage{diagbox}

\newtcbox{\mymath}[1][]{%
    nobeforeafter, math upper, tcbox raise base,
    enhanced, colframe=blue,
    colback=yellow!30, boxrule=1pt,
    #1}
    
\usepackage{natbib}
\usepackage{draftwatermark}
\SetWatermarkText{}
\SetWatermarkScale{4}

\newtheorem{theorem}{Theorem}

\newtheorem{claim}{Claim}

\newtheorem{example}{Example}

\newenvironment{definition}[1][Definition]{\noindent\textbf{#1.} }{}
\usepackage{tikz}
\renewcommand{\emptyset}{\varnothing}

\begin{document}
\title{Identification in the Random Utility Model}
\thanks{I thank the editor, Faruk Gul, and two anonymous referees for their comments that have helped improve this paper. I also thank Luca Anderlini, Axel Anderson, Roger Lagunoff, Jay Lu, Marco Mariotti, Alexandre Poirier, Collin B. Raymond, John Rehbeck, Tomasz Strzalecki, and seminar participants at D-TEA 2021 for helpful comments and discussions. I am especially grateful to Peter Caradonna, Christopher Chambers, and Yusufcan Masatlioglu for their continued support and insightful conversations throughout the course of this project.\\
Turansick:  Department of Economics, Georgetown University, ICC 580  37th and O Streets NW, Washington DC 20057.  E-mail:  \texttt{cmt152@georgetown.edu} }
\author{Christopher Turansick}
\date{\today}
\maketitle

\begin{abstract}
The random utility model is known to be unidentified, but there are times when the model admits a unique representation. We offer two characterizations for the existence of a unique random utility representation. Our first characterization puts conditions on a graphical representation of the data set. Non-uniqueness arises when multiple inflows can be assigned to multiple outflows on this graph. Our second characterization provides a direct test for uniqueness given a random utility representation. We also show that the support of a random utility representation is identified if and only if the representation itself is identified.
    
\end{abstract}



\pagebreak

\section{Introduction}
The fundamental goal of revealed preference theory is recovering an agent's preference from choice data. When faced with rational agents, revealed preference theorists have shown that recovery can be achieved. However, agents are not rational as choices are stochastic. The natural relaxation of the standard rationality assumption is to stochastic rationality; agents are rational conditional on a varying unobserved state. This type of rationality is modeled by the random utility model of \citet{block1959random}. Instead of a single preference, agents possess a distribution over preferences. The goal now is to recover a distribution over preferences instead of just a single preference. Unfortunately, there has been less success in recovering a distribution of preferences.

While the random utility model is identified when there are three or fewer alternatives \citep{block1959random}, for larger environments, the random utility model is in general not identified \citep{barbera1986falmagne,fishburn1998stochastic}.\footnote{\citet{falmagne1978representation}, \citet{barbera1986falmagne}, and \citet{gibbard2021disentangling} tell us that the random utility model is identified up to the probability weights on contour sets.} The heart of the identification problem is as follows. We can recover the probability that $w$ is preferred to $x$ and the probability that $y$ is preferred to $z$, but we cannot necessarily recover the probability that both $w$ is preferred to $x$ and $y$ is preferred to $z$ \citep{strzalecki2017stochastic}. Notably, not even the support of the rationalizing distribution is guaranteed to be identified. This means that analysts are unable to even recover the types of preferences in a population. There are distributions with disjoint supports that induce the same set of choice probabilities \citep{fishburn1998stochastic}.

This uniqueness problem gives rise to both empirical and theoretical concerns. From a theoretical perspective, identification of a model allows theorists to map the parameters of their models to behavioral outcomes. One of the main goals of choice theory is to provide simplified approximations of reality in an attempt to explain observed choice behavior. Identification of a model allows us to do exactly this. From an empirical perspective, identification of a choice model allows social planners and mechanism designers to perform proper counterfactual analysis. Identification guarantees that counterfactual analysis will be accurate up to the choice of model. When choice behavior has multiple representations, counterfactuals may take on different values for each one of these representations. This is especially important when we are considering counterfactuals and policy questions that rely on more than just choice frequencies.

To accommodate the random utility model's lack of identification, we ask a new type of question. We ask which rationalizable data sets can be uniquely represented by the random utility model. This differs from the standard approach in the literature which puts further assumptions on the random utility model to ensure that every rationalizable data set is uniquely represented. We take our approach primarily for two reasons. Our first reason is to avoid the restrictive behavioral implications of identifying assumptions. As an example, consider the Luce model and its variants. In practice, these are the most commonly used random utility models. Identification of these models comes at the cost of behavioral implications that are not observed in practice and ex ante unreasonable counterfactuals.\footnote{Behavioral and counterfactual shortcomings of the Luce class of models are well documented. To name a few, we have the red bus/blue bus problem \citep{debreu1960review}, overestimation of demand for goods with high prices \citep{bajari2001discrete}, misleading cross elasticities \citep{ackerberg2005unobserved}, and demand being discontinuous in characteristics \citep{lu2021mixed}.} Our second reason for this approach is to give insight in to how non-uniqueness arises. As part of our analysis, we pin down the exact graphical structure of non-uniqueness in the random utility model. We believe this will aid researchers in developing new identified random utility models.\footnote{Notably, \citet{fishburn1998stochastic} develops an example that shows one way that non-uniqueness arises. We show that, in essence, this is the only way that non-uniqueness arises.}

In order to pin down the graphical structure of non-uniqueness, we use the graphical construction of \citet{fiorini2004short} to take a look at the counterexample of \citet{fishburn1998stochastic}. Our first result generalizes the structure of this example and characterizes which data sets have a unique random utility representation. Non-uniqueness arises when multiple inflows can be assigned to multiple outflows. It is the fact that the random utility model does not pin down the assignment of inflows to outflows that causes non-uniqueness. Any identified random utility model will either preclude the event where this assignment problem arises or pin down the assignment of inflows to outflows.

Our second result characterizes which distributions over preferences are observationally unique. To do this, we translate the conditions used in our first result into conditions about the contour sets of preferences in the support of the distribution. A distribution will be observationally equivalent to some other distribution if there are two preferences in the support that satisfy the following. The two preferences must share some common upper and associated lower contour set and the ordering within these two contour sets differ between the preferences. For larger choice environments, this result means that a unique representation and a full support representation are mutually exclusive.\footnote{This observation was first noted in \citet{mcclellon2015unique}. The result of \citet{mcclellon2015unique} speaks to data sets that are generated by full support distributions. Our result does not require knowledge of the generating process. We only need knowledge of observables.}

The last of our three results characterizes when the support of a random utility representation is uniquely identified. With complete data, every random utility representation has the same support if and only if the random utility representation is unique. In other words, when we have complete data, pinning down the support of a representation is just as hard as pinning down the representation itself.

The rest of this paper is organized as follows. We close this section with a review of related literature. Section 2 reviews the random utility model and the counterexample to uniqueness of \citet{fishburn1998stochastic}. In Section 3, we introduce and discuss our main results. We conclude in Section 4.

\subsection{Related Literature}
Our paper builds on the literature that studies the empirical content of the random utility model of \citet{block1959random}. \citet{falmagne1978representation}, \citet{barbera1986falmagne}, \citet{mcfadden1990stochastic}, and \citet{fiorini2004short} offer characterizations and discussions of the random utility model. More closely related to our work, there is a strand of literature that studies the uniqueness properties of the random utility model. \citet{fishburn1998stochastic} offers an example which shows that the random utility model is not identified. \citet{mcclellon2015unique} uses and extends this example to show that the problem of non-uniqueness is widespread for larger choice environments. \citet{dardanoni2020mixture} and \citet{gibbard2021disentangling} study random utility when agents have limited cognitive ability. \citet{dardanoni2020mixture} study identification of both the underlying preferences as well as the cognitive parameters of the decision makers using a stronger type of data, mixture choice data. \citet{gibbard2021disentangling} studies which uniqueness properties of the random utility model remain when agents have limited attention. Our paper lies in the intersection of these two strands of literature. We offer a characterization of when the random utility model admits a unique representation.

Our paper is also related to the literature which extends the random utility model in order to recover uniqueness. \citet{gul2006random} extend the random utility model to choice over lotteries. They show that restricting the set of preferences to expected utility preferences recovers uniqueness. \citet{lin2020random} studies this uniqueness result by considering to what extent the axioms of expected utility can be relaxed while maintaining uniqueness. Lin shows that the relaxation of the independence axiom to the betweenness axiom \citep{dekel1986axiomatic} causes uniqueness to be lost. \citet{yang2021random} considers randomization over quasi-linear preferences and choice over price-indexed bundles. The restriction to quasi-linear preferences leads to identification of the model. These papers recover uniqueness by restricting the set of preferences allowed by the model.

More recently, there is a strand of literature which recovers uniqueness by putting assumptions on the support of a random utility representation.  \citet{apesteguia2017single} are the first to do this. They extend the random utility model by asking that the support of the representation satisfy the single-crossing property with respect to an exogenous order. \citet{filiz2020progressive} extend this result by considering randomization over choice functions while maintaining the single-crossing assumption. \citet{honda2021random} takes a different approach. Instead of assuming single-crossing, Honda assumes a random cravings condition. The random cravings condition supposes that there is some underlying true preference and that every preference in the support of a representation only differs from this true preference by the ranking of a single alternative. Unlike the prior collection of papers, the models of these papers allow any preference to be in the support of some representation. These papers simply restrict which preferences can concurrently be in the support of a single representation.

\section{The Random Utility Model and Fishburn}
To begin, we review the random utility model (RUM). Let $X$ be a finite set of alternatives. Let $\Pi$ be the set of linear orders over $X$.\footnote{A linear order is an antisymmetric, transitive, and complete binary relation.} We use $\pi$ to denote an element of $\Pi$. We use the notation $\pi(A)>\pi(B)$ to denote that every element of $A$ is ranked higher than every element of $B$ according to $\pi$. This implies no further restrictions on how $\pi$ ranks elements of $A$ against other elements of $A$. The same is true for elements of $B$. When $A = \{x\}$, we use the notation $\pi(x)$. Further, we call $\Delta (\Pi)$ the set of probability distributions over $\Pi$. We say that an agent makes decisions according to RUM if they are endowed with a $\nu \in \Delta (\Pi)$ and, whenever they make a decision, they draw a linear order according to this $\nu$ and then choose the maximal element according to the drawn linear order.

We consider stochastic choice data for each non-empty subset of $X$. To formalize this, the data we consider is called a system of choice probabilities. A pair $(X,P)$ is a system of choice probabilities if for all non-empty subsets $A$ of $X$, $P_A(\cdot)$ defines a probability distribution over the elements of $A$. A system of choice probabilities captures the choice probability of each element $x$ of each non-empty subset $A$ of $X$. We now define what it means for data to be rationalizable by RUM.

\begin{definition}
We say that a system of choice probabilities is rationalizable if there exists some $\nu \in \Delta (\Pi)$ such that for all non-empty $A \subseteq X$ and all $x \in A$ we have
$$P_A(x) = \sum_{\pi \in \Pi} \nu( \pi) \mathbf{1}\{ \pi(x) > \pi(A \setminus \{x\}) \}.$$
\end{definition}

\citet{falmagne1978representation} was the first to characterize rationalizability for RUM. The characterization relies on the Block-Marschak polynomials, henceforth BM-polynomials, which were first introduced by \citet{block1959random}. We state the definition of the BM-polynomials here.

\begin{definition}
For a non-empty set $A \subseteq X$ and an element $x \in A$, the BM-polynomial for $x$ in $A$ is given by
\begin{equation*}
    \begin{split}
        q(x,A) & = P_A(x) - \sum_{A \subsetneq A'} q(x,A') \\
        & = \sum_{A \subseteq A'} (-1)^{|A' \setminus A|} P_{A'}(x).
    \end{split}
\end{equation*}
\end{definition}

To interpret the BM-polynomials, we turn to a result of \citet{falmagne1978representation}. Let $M_{x,A}$ be the set of linear orders on $X$ that rank $x$ exactly at the top of $A$.
$$M_{x,A}=\{\pi|\pi(X \setminus A) > \pi(x) > \pi(A \setminus \{x\}) \}$$
\citet{falmagne1978representation} shows that a distribution $\nu$ rationalizes a system of choice probabilities if and only if $q(x,A) = \nu(M_{x,A})$ for all all such $x \in A \subseteq X$. This is also the classic uniqueness result for RUM. Any two representations of a system of choice probabilities must put the same probability weight on each contour set. The characterization of RUM by \citet{falmagne1978representation} states that all BM-polynomials must be non-negative. We are interested in characterizing when the rationalizing probability distribution is unique.

Our characterization combines the graphical representation of RUM presented in \citet{fiorini2004short} with the intuition of the counterexample to uniqueness presented in \citet{fishburn1998stochastic}. We begin with the graphical construction due to \citet{fiorini2004short}. Consider a graph with nodes indexed by the elements of $2^X$, the power set of $X$. We will use the set indexing a node to refer to that node. There exists an edge between two nodes $A$ and $B$ if one of the following is true.
\begin{enumerate}
    \item $A \subseteq B$ and $|B \setminus A| = 1$
    \item $B \subseteq A$ and $|A \setminus B| = 1$
\end{enumerate}
In other words, the edge set of this graph is formed by applying the covering relation of $\subseteq$ to $X$. Now we assign weights to these edges. Assign $q(x,A)$ to the edge connecting $A$ and $A\setminus \{x\}$. \citet{fiorini2004short} does not give a name to this graph, but we will refer to it as the \textbf{probability flow diagram}. Figure 1 gives an example of the probability flow diagram for the set $X = \{a,b,c\}$.
\begin{figure}
    \centering
\begin{tikzpicture}[scale=.5, transform shape]
    \tikzstyle{every node} = [rectangle]
    
        \node (a) at (6,0) {$\emptyset$};
        
        \node (b) at (0,5) {$\{a\}$};
        \node (c) at (6,5) {$\{b\}$};
        \node (d) at (12,5) {$\{c\}$};
        
        \node (e) at (0,10) {$\{a,b\} $};
        \node (f) at (6,10) {$\{a,c\}$};
        \node (g) at (12,10) {$\{b,c\}$};

        \node (h) at (6,15) {$\{a,b,c\}$};
        
        \draw [->] (h) -- (f) node[midway, below, sloped] {$q(b,\{a,b,c\})$};
        \draw [->] (h) -- (g) node[midway, below, sloped] {$q(a,\{a,b,c\})$};  
        \draw [->] (h) -- (e) node[midway, below, sloped] {$q(c,\{a,b,c\})$};

        \draw [->] (e) -- (b) node[midway, below, sloped] {$q(b,\{a,b\})$}; 
        \draw [->] (e) -- (c) node[pos=.25, below, sloped] {$q(a,\{a,b\})$};
        \draw [->] (f) -- (b) node[pos=.25, below, sloped] {$q(c,\{a,c\})$};
        \draw [->] (f) -- (d) node[pos=.25, below, sloped] {$q(a,\{a,c\})$};
        \draw [->] (g) -- (c) node[pos=.25, below, sloped] {$q(c,\{b,c\})$};
        \draw [->] (g) -- (d) node[midway, below, sloped] {$q(b,\{b,c\})$}; 

        \draw [->] (b) -- (a) node[midway, below, sloped] {$q(a,\{a\})$};
        \draw [->] (c) -- (a) node[midway, below, sloped] {$q(b,\{b\})$};
        \draw [->] (d) -- (a) node[midway, below, sloped] {$q(c,\{c\})$};
    
    \end{tikzpicture}
    \caption{The probability flow diagram for the set $X = \{a,b,c\}$.}
    \label{fig:probflow}
\end{figure}
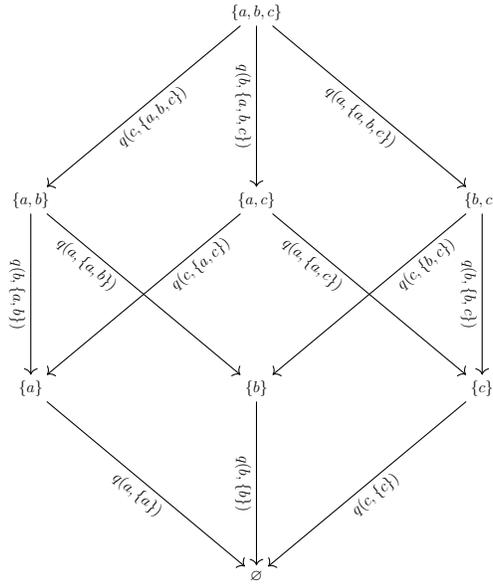

We now revisit the counterexample of \citet{fishburn1998stochastic} and explore the probability flow diagram of the counterexample.
\begin{example}[Fishburn's Counterexample]
Let $X = \{a,b,c,d\}$. Consider the following probability distributions over linear orders on $X$.
\begin{equation*}
    \nu_1(\pi) =    \begin{cases}
                    \frac{1}{2} & \text{if }\pi \in \{ a \succ b \succ c \succ d, b \succ a \succ d \succ c\} \\
                    0 & \text{otherwise}
                 \end{cases}
\end{equation*}
\begin{equation*}
    \nu_2(\pi) =    \begin{cases}
                    \frac{1}{2} & \text{if }\pi \in \{ a \succ b \succ d \succ c, b \succ a \succ c \succ d\} \\
                    0 & \text{otherwise}
                 \end{cases}
\end{equation*}
These two probability distributions induce the same system of choice probabilities.
\end{example}

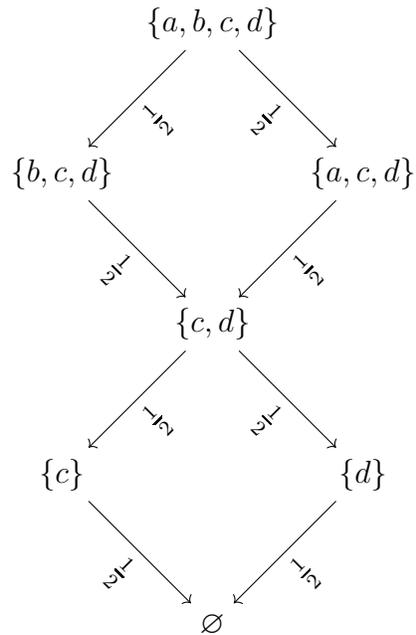
\begin{figure}
    \centering
    \begin{tikzpicture}[scale=.5]
    \tikzstyle{every node} = [rectangle]
        \node (a) at (8,0) {$\emptyset$};

        \node (b) at (4,4) {$\{c\}$};
        \node (c) at (12,4) {$\{d\}$};
        
        \node (d) at (8,8) {$\{c,d\}$};
        
        \node (e) at (4,12) {$\{b,c,d\}$};
        \node (f) at (12,12) {$\{a,c,d\}$};
        
        \node (g) at (8,16) {$\{a,b,c,d\}$};
        
        \draw [->] (g) -- (e) node[midway, below, sloped] {$ \frac{1}{2}$};
        \draw [->] (g) -- (f) node[midway, below, sloped] {$ \frac{1}{2}$};
        
        \draw [->] (e) -- (d) node[midway, below, sloped] {$ \frac{1}{2}$};
        \draw [->] (f) -- (d) node[midway, below, sloped] {$ \frac{1}{2}$};

        \draw [->] (d) -- (b) node[midway, below, sloped] {$ \frac{1}{2}$};
        \draw [->] (d) -- (c) node[midway, below, sloped] {$ \frac{1}{2}$};

        \draw [->] (b) -- (a) node[midway, below, sloped] {$ \frac{1}{2}$};
        \draw [->] (c) -- (a) node[midway, below, sloped] {$ \frac{1}{2}$};
        
    \end{tikzpicture}
    \caption{The reduced probability flow diagram for the Fishburn counterexample.}
    \label{fig:FishEx}
\end{figure}
This is the counterexample which \citet{fishburn1998stochastic} uses to show that RUM is not identified. Figure 2 shows the reduced probability flow diagram of this example.\footnote{By reduced probability flow diagram we mean that we take the probability flow diagram and remove each edge with zero weight and each node whose connected edges all have zero weight.} In the example above, as both probability distributions induce the same system of choice probabilities, they have the same probability flow diagram. The key feature of this example is found at the node $\{c,d\}$. Note that there are two edges with strictly positive weight that go into $\{c,d\}$ and two edges with strictly positive weight that leave $\{c,d\}$. It turns out that this two-in and two-out structure exactly characterizes non-uniqueness in RUM. We generalize this two-in and two-out structure in the next section.

\section{Characterizing Uniqueness}
In this section, we present two characterizations which tell us when a RUM representation is unique. The first characterization tests choice data while the second tests the representation itself. Using these results, we then characterize when the support of a rationalizing distribution is unique. We now introduce the terminology needed to state our characterizations.

\begin{definition}
We call a path $\rho$ a finite sequence of sets $\{A_i\}_{i=0}^{|X|}$ such that $A_{i+1}\subsetneq A_i$ for all $i$, $A_0=X$, and $A_{|X|}=\emptyset$.
\end{definition}

\citet{fiorini2004short} notes that there is a bijection between paths on the probability flow diagram and the set of linear orders of $X$. The bijection pairs the path $\{X,X\setminus \{x_1\},X \setminus \{x_1,x_2\}, \dots ,\emptyset\}$ with the order that ranks $x_1\succ x_2 \succ \dots$. When we construct a representation, the probability weight associated with order $\pi$ is derived as follows. We decompose the probability flow diagram into paths flows. We then assign the path flow of the path corresponding to $\pi$ as the probability weight put on $\pi$. We will be using the prior bijection and the associated edge weights to study which orders can receive a strictly positive weight in a representation. This idea is captured graphically by the following definition.

\begin{definition}
For a system of choice probabilities $(X,P)$ and its corresponding probability flow diagram, we call a path supported if for all $i \in \{0, \dots, |X|-1\}$, $q(A_i \setminus A_{i+1},A_i)>0$.
\end{definition}

There exists a representation which puts strictly positive weight on a linear order $\pi$ if and only if the path associated with $\pi$ is supported.\footnote{To see this, note that if the algorithm used in the proof of Theorem 1 begins by subtracting out the considered supported path, then that linear order $\pi$ is in the support of the representation. Further, any order which has a path which is not supported must necessarily receive zero probability weight in a representation.} Due to this, if a system of choice probabilities has multiple representations, it must be that the differing probability weights are restricted to orders which have supported paths. As we mentioned prior, the characterization for uniqueness relies on the idea of two-in and two-out. The definition of branching formalizes this idea.

\begin{definition}
We call two paths $\rho$ and $\rho'$ branching if there exists some $i\leq j$ with $i,j \in \{1,\dots,|X|-1\}$ such that $A_{i-1}^{\rho} \neq A_{i-1}^{\rho'}$, $A_{j+1}^{\rho} \neq A_{j+1}^{\rho'}$, and for all $m \in \{i, \dots, j\}$, $A_m^{\rho} = A_m^{\rho'}$.
\end{definition}

Unlike in the counterexample of \citet{fishburn1998stochastic}, the definition of branching does not require the two-in and two-out to happen at the same node. The definition of branching allows for two paths to go into the same node, share a few common edges, and then split. We now have all the terminology we need to state our first theorem.

\begin{theorem}
Suppose that a system of choice probabilities $(X,P)$ is rationalizable. The rationalizing $\nu$ is unique if and only if the probability flow diagram has no pairs of supported branching paths.
\end{theorem}

We leave all proofs to the appendix. However, we discuss the intuition of the proof here. To see the logic for necessity, first consider a node that satisfies two-in and two-out. Call the two-in edges $a$ and $b$ respectively. Call the two-out edges $c$ and $d$ respectively. We can construct two disjoint sets of paths that induce this two-in and two-out property. Consider the pair of paths $\{(a,c),(b,d)\}$. These two paths satisfy two-in and two-out at the considered node. Similarly, the pair of paths $\{(a,d),(b,c)\}$ satisfy two-in and two-out along the same edges as the first pair of paths. This shows that two supported branching paths imply non-uniqueness.

To see the logic for sufficiency, we first note that if no pair of supported paths satisfy two-in and two-out, then every supported path that satisfies two-out with some other supported path must do so above any node at which they satisfy two-in. Similarly, any supported path that satisfies two-in with some other supported path must do so below any node at which they satisfy two-out. These two facts mean that for every supported path there exists some edge such that any two-in happens below that edge and every two-out happens above that edge. The weight along this edge uniquely identifies the probability weight put on the order associated with this path.

Note that Theorem 1 subsumes some known results. \citet{block1959random} tells us that, when $|X| \leq 3$, any representation is unique. We note that branching paths are not found unless $|X| \geq 4$.\footnote{To see this, observe the following. A pair of branching paths share the node $X$, have differing nodes somewhere below $X$, have a common node below their differing nodes, have another differing node below their common node, and then share the node $\emptyset$. This requires having five nodes which can only happen when $|X| \geq 4$.} As an immediate corollary of Theorem 1, we are able to show that, when $|X| \leq 3$, any representation of a system of choice probabilities is unique.

\citet{mcclellon2015unique} shows that when $|X| \geq 4$, the issue of non-uniqueness is widespread. The result notes that if a system of choice probabilities is induced by a full support distribution, then there is a different distribution which induces the same system of choice probabilities. If a system of choice probabilities is induced by a full support distribution, then every path is supported. Since there are branching paths when $|X| \geq 4$, it follows immediately that any representation is not unique. This result can be extended to say that if every BM-polynomial of a system of choice probabilities is strictly positive, then the representation is not unique.

This result has an interesting implication for the Luce and logit class of models. Since Luce choice probabilities can be represented by the logit model and the logit model induces a full support over preferences, for $|X|\geq 4$, any random utility representation of any system of choice probabilities consistent with Luce is not unique. This observation extends to other statistical discrete choice models with full support including the generalized extreme values model (see \citet{McFadden1977ModellingTC} and \citet{dagsvik1995large}).

We now move onto our second characterization. Intuitively, this characterization takes the structure of branching paths and restates that structure in terms of properties of the contour sets of the associated orders. Checking that the support of a representation satisfies these properties amounts to a finite test. Before moving forward, we state the definition of upper contour set.

\begin{definition}
The weak upper contour set of some $x \in X$ according to $\pi \in \Pi$ is the set of all elements $y \in X$ such that $\pi(y) \geq \pi(x)$. We write
$$U_{\pi}(x) = \{y | \pi(y) \geq \pi(x)\}$$
to denote the weak upper contour set of $x$ according to $\pi$.
\end{definition}

With this definition, we are now able to state our second characterization.

\begin{theorem}
Suppose that a system of choice probabilities $(X,P)$ is rationalizable. The rationalizing $\nu$ is unique if and only if there are no pairs of orders $\pi$ and $\pi'$ satisfying the following.
\begin{enumerate}
    \item $\nu(\pi)>0$ and $\nu(\pi')>0$
    \item There exists $x,y,z \in X$ such that
    \begin{enumerate}
    \item $\pi(\{x,y\}) > \pi(z)$ and $\pi'(\{x,y\}) > \pi'(z)$
    \item $x \neq y$
    \item $U_{\pi}(z) \neq U_{\pi'}(z)$
    \item $U_{\pi}(x) = U_{\pi'}(y)$
\end{enumerate}
\end{enumerate}
\end{theorem}

Intuitively, the first condition of Theorem 2 captures the definition of a supported path and the second condition captures the definition of a pair of branching paths. Our proof consists of showing that the existence of a pair of supported branching paths is equivalent to the two conditions of Theorem 2. Necessity follows primarily from definitions. The logic for sufficiency is as follows. We first suppose that the representation is not unique. Then, by Theorem 1, there must be a pair of supported branching paths. We show that no matter how one allocates the weight from these supported branching paths, there will always be two orders which violate the conditions of Theorem 2. Now, consider the following definition.

\begin{definition}
Let $S_{\nu}= \{\pi|\nu(\pi) >0 \}$ be the set of linear orders with strictly positive weight under representation $\nu$. Then we call $S_\nu$ the support of $\nu$.
\end{definition}

Suppose we are in the case where there is non-uniqueness. Theorem 2 tells us that there are two linear orders in the support of our representation that have different rankings of $x$ and $y$, rank both above $z$, and differ in their ranking of $z$. This means that there is some fourth element $w$ such that $w$ is ranked below $x$ and $y$ in both orders and is ranked differently compared to $z$ in both rankings. If we restrict to the choice problem of $X=\{w,x,y,z\}$, this is exactly the example of \citet{fishburn1998stochastic} with potentially differing probability weights. A consequence of this observation is that if the system of choice probabilities $(X,P)$ is not uniquely represented, then there exists some subset of $X$ of size four, call it $Y \subseteq X$, such that $(Y,P)$ is not uniquely represented. In other words, if an analyst wants to check for uniqueness, it is sufficient to check for uniqueness of each system of choice probabilities induced by sets of size four. This further means that, subject to finding a representation, if we observe choices only from sets of size four and smaller and we have a unique representation, we can recover the choice probabilities of sets larger than size four. This is not true in the general case where observing choices from larger choice sets further restricts the set of potential representations.

Another interesting consequence of Theorem 2 is that uniqueness is a property of the support of a rationalizing distribution. This means that if a distribution over preferences is the unique representation for some system of choice probabilities, then any distribution with the same support is also the unique representation for its system of choice probabilities. The opposite is also true. If a distribution over preferences is observationally equivalent to some other distribution over preferences, then any distribution over preferences with the same support is observationally equivalent to some other distribution over preferences. It turns out that uniqueness of a representation is equivalent to the support of a representation being identified.

\begin{theorem}
Suppose that a system of choice probabilities $(X,P)$ is rationalizable. Each rationalizing distribution has the same support if and only if $(X,P)$ is uniquely rationalizable.
\end{theorem}

Obviously, if the support is not identified then the rationalizing distribution will not be identified. The intuition for the other direction of the proof is much the same as the intuition for Theorem 1. If the rationalizing distribution is not identified, then the probability flow diagram will have the two-in and two-out structure. Just as before, this means we can decompose the two-in and two-out structure into two disjoint sets of paths. These two disjoint sets then represent two disjoint sets of preferences which induce the two-in and two-out structure. This shows that non-uniqueness of the distribution of preferences implies non-uniqueness of the support of preferences.

\section{Discussion}
In this paper, we provide two characterizations for when a random utility representation is unique. Theorem 1 provides conditions on the BM-polynomials which characterize the graphical structure of non-uniqueness. As BM-polynomials are rarely used in empirical settings, we think of Theorem 1 as a theoretical tool for developing other identified random utility models. Theorem 2 gives conditions which tell us when a representation is observationally equivalent to some other representation. We view Theorem 2 as being a potential empirical tool. Once a representation is found using standard methods (see \citet{kitamura2018nonparametric} and \citet{smeulders2021nonparametric}), Theorem 2 can be applied to check for uniqueness.

As an application of our results, we now consider the single-crossing random utility model (SCRUM) of \citet{apesteguia2017single}. SCRUM puts an additional restriction on the underlying structure of $X$ in that $X$ is endowed with some exogenous linear order $\rhd$. We say a system of choice probabilities is rationalizable by SCRUM if there exists some RUM representation of the system, $\nu$, such that the support of $\nu$ can be ordered so that it satisfies the single-crossing property with respect to $\rhd$. Recall that the single-crossing property is as follows.

\begin{definition}
We say that a representation $\nu$ satisfies the single-crossing property if the support of $\nu$ can be ordered in such a way that for all $x \rhd y$, $\pi_i(x) > \pi_i(y)$ implies $\pi_j(x) > \pi_j(y)$ for all $j \geq i$.
\end{definition}

\citet{apesteguia2017single} show that a SCRUM representation is unique. We now return to the counterexample of \citet{fishburn1998stochastic}.

\begin{example}[Fishburn and SCRUM]
Let $X = \{a,b,c,d\}$. Consider the system of choice probabilities induced by the following two distributions over linear orders.
\begin{equation*}
    \nu_1(\pi) =    \begin{cases}
                    \frac{1}{2} & \text{if }\pi \in \{ a \succ b \succ c \succ d, b \succ a \succ d \succ c\} \\
                    0 & \text{otherwise}
                 \end{cases}
\end{equation*}
\begin{equation*}
    \nu_2(\pi) =    \begin{cases}
                    \frac{1}{2} & \text{if }\pi \in \{ a \succ b \succ d \succ c, b \succ a \succ c \succ d\} \\
                    0 & \text{otherwise}
                 \end{cases}
\end{equation*}
Suppose the exogenous order on $X$ is $a \rhd b \rhd c \rhd d$. Then the system of choice probabilities is SCRUM rationalized by $\nu_1$. Now suppose that the exogenous order on $X$ is $a \rhd b \rhd d \rhd c$. Then the system of choice probabilities is SCRUM rationalized by $\nu_2$.

\end{example}

Recall that when a system of choice probabilities has multiple representations, it essentially embeds the example of \citet{fishburn1998stochastic} in some subset of size four. As we see in the above example, the SCRUM representation of the Fishburn example is pinned down by the exogenous order $\rhd$. This follows from the fact that if $\nu_1$ satisfies the single-crossing property with respect to $\rhd$ it must be the case that $\nu_2$ does not. Uniqueness in SCRUM now follows from extensions of this logic.

We see three potential extensions of our work. In this paper, we have maintained the assumption that we observe choice on every non-empty subset of $X$. In empirical settings, this is often unreasonable. One natural extension of our work is to consider the same question but with choice on a limited domain. A potential second extension is to generalize our results to infinite choice domains. This extension would provide insight for model builders who consider choices over lotteries \citep{gul2006random}, dynamic choice \citep{frick2019dynamic}, or choice over price-indexed bundles \citep{yang2021random}.

The third potential extension we consider utilizes the algorithmic nature of our approach. Instead of asking when we have a unique representation, one may be interested in recovering the set of representations for a given system of choice probabilities. As the set of representations for a given system of choice probabilities is convex, this amounts to finding the extreme points of the set of representations. In our construction of a representation, we consider a specific order in which we decompose the probability flow diagram into path flows (and thus assign probability weights to the corresponding order). However, one could exogenously vary the order in which path flows are subtracted from the probability flow diagram. Consider the following description of an algorithm.
\begin{enumerate}
    \item Choose some yet unchosen order over paths.
    \item Decompose the probability flow diagram into path flows according to the chosen order. Assign the smallest remaining edge capacity of the considered path as the path flow of that path. (See the proof of Theorem 1.)
    \item Repeat the prior steps until every order over paths has been exhausted.
\end{enumerate}
This algorithm will return a collection of representations, each of which is an extreme point of the set of representations.\footnote{To see this, note that every distribution found by this algorithm can be matched with an order/ranking over linear orders. The representation associated with a specific ranking over linear orders places the most probability weight possible on a given linear order conditional on higher ranked linear orders getting the most probability weight possible with further iterative conditioning.} It is an open question whether this algorithm returns every extreme point of the set of representations.

At the end of the day, identified models are appealing as they allow for proper counterfactual analysis and clean interpretation of parameters. Our main insight tells us when we can treat the random utility model as if it were identified. This insight offers aid in the future construction of identified random utility models.

\singlespacing
\appendix 
\section{Omitted Proofs}

\begin{definition}
We call two paths $\rho$ and $\rho'$ in-branching if there exists some $i \in \{1,\dots, |X|-1\}$ such that $A_i^{\rho} = A_i^{\rho'}$ and $A_{i-1}^{\rho} \neq A_{i-1}^{\rho'}$
\end{definition}

\begin{definition}
We call two paths $\rho$ and $\rho'$ out-branching if there exists some $i \in \{1,\dots, |X|-1\}$ such that $A_i^{\rho} = A_i^{\rho'}$ and $A_{i+1}^{\rho} \neq A_{i+1}^{\rho'}$
\end{definition}

\begin{definition}
We call a collection of sets, $\{A_i, \dots, A_j\}$, a branching section of paths $\rho$ and $\rho'$ if $A_{i-1}^{\rho} \neq A_{i-1}^{\rho'}$, $A_{j+1}^{\rho} \neq A_{j+1}^{\rho'}$, and for all $m \in \{i, \dots, j\}$, $A_m^{\rho} = A_m^{\rho'}$.
\end{definition}

\subsection{Proof of Theorem 1}

We begin by showing necessity. We proceed by contraposition. Suppose there are two supported paths, $\rho$ and $\rho'$, that are branching. This means that these two paths share some set of common nodes $\{A_n,\dots,A_m\}$ such that $A_{n-1}^{\rho} \neq A_{n-1}^{\rho'}$ and $A_{m+1}^{\rho} \neq A_{m+1}^{\rho'}$. Consider the following two paths, respectively $\rho''$ and $\rho'''$, $(A_0^{\rho}, \dots, A_m^{\rho},A_{m+1}^{\rho'},\dots,A_{|X|}^{\rho'})$ and $(A_0^{\rho'}, \dots, A_m^{\rho'},A_{m+1}^{\rho},\dots,A_{|X|}^{\rho})$. Note that the node set and the edge set of $\rho \cup \rho'$ are the same as the node set and edge set of $\rho'' \cup \rho'''$. Let $r$ be the minimum flow along the edge set of $\rho \cup \rho'$. Without loss, let $r$ be the flow of an edge that belongs to the edge set of $\rho$ and $\rho''$. We will now construct two different representations. We construct $\nu_1$ as follows.
\begin{enumerate}
    \item Let $\nu_1(\pi_{\rho})=r$.
    \item For all $q(\cdot,\cdot)$ on the edge set of $\rho$, let $q_0(\cdot,\cdot)=q(\cdot,\cdot)-r$. For all $q(\cdot,\cdot)$ not on the edge set of $\rho$, let $q_0(\cdot,\cdot)=q(\cdot,\cdot)$.
    \item Initialize at $i=0$.
    \item Let $s$ be the smallest strictly positive $q_i(\cdot,\cdot)$. Choose some edge which has flow equal to $s$. Since inflow equals outflow (see explanation below the algorithms), this edge is a part of some path from $X$ to $\emptyset$ with all edges along the path having strictly positive flow. Fix this path and call it $\gamma$.
    \item Let $\pi_i$ denote the linear order that is bijectively associated with $\gamma$. Set $\nu_1(\pi_i)=s$.
    \item For all edges along path $\gamma$, let $q_{i+1}(\cdot,\cdot) = q_i(\cdot,\cdot) -s$. For all edges not along path $\gamma$, let $q_{i+1}(\cdot,\cdot) = q_i(\cdot,\cdot)$.
    \item If there is strictly positive flow anywhere along the graph, return to step 4. If not, terminate the algorithm.
\end{enumerate}
We construct $\nu_2$ as follows.
\begin{enumerate}
    \item Let $\nu_2(\pi_{\rho''})=r$.
    \item For all $q(\cdot,\cdot)$ on the edge set of $\rho''$, let $q_0(\cdot,\cdot)=q(\cdot,\cdot)-r$. For all $q(\cdot,\cdot)$ not on the edge set of $\rho''$, let $q_0(\cdot,\cdot)=q(\cdot,\cdot)$.
    \item Initialize at $i=0$.
    \item Let $s$ be the smallest strictly positive $q_i(\cdot,\cdot)$. Choose some edge which has flow equal to $s$. Since inflow equals outflow, this edge is a part of some path from $X$ to $\emptyset$ with all edges along the path having strictly positive flow. Fix this path and call it $\gamma$.
    \item Let $\pi_i$ denote the linear order that is bijectively associated with $\gamma$. Set $\nu_2(\pi_i)=s$.
    \item For all edges along path $\gamma$, let $q_{i+1}(\cdot,\cdot) = q_i(\cdot,\cdot) -s$. For all edges not along path $\gamma$, let $q_{i+1}(\cdot,\cdot) = q_i(\cdot,\cdot)$.
    \item If there is strictly positive flow anywhere along the graph, return to step 4. If not, terminate the algorithm.
\end{enumerate}
Note that we know from \citet{fiorini2004short} and \citet{falmagne1978representation} that we have inflow equals outflow on this graph at the start of each of these algorithms. Since each iteration of the algorithm subtracts out a fixed amount from each edge of a given path, we have inflow equals outflow at each stage of this algorithm. This means that this algorithm terminates with zero flow along the graph. To see this, suppose not. Then there is positive flow somewhere along the graph at termination. Since we have inflow equals outflow, we can follow this positive flow all the way to the nodes $X$ and $\emptyset$. This then shows that there is some path with strictly positive flow, thus contradicting termination of our algorithm. Further, this algorithm assigns $q(x,A)$ to orders that rank $x$ exactly at the top of $A$. Thus, we know from \citet{falmagne1978representation} that $\nu_1$ and $\nu_2$ rationalize the system of choice probabilities. Now note that since there is an edge that is shared between $\rho$ and $\rho''$ which has flow equal to $r$, $\nu_1$ puts zero weight on $\pi_{\rho''}$ while $\nu_2$ puts weight equal to $r$ on $\pi_{\rho''}$. Thus these two representations are different, meaning that there is not a unique representation. By contraposition, we have proven necessity.

Now we prove sufficiency. Suppose no two supported paths are branching.
\begin{claim}
Let $\rho$ and $\rho'$ be supported out-branching paths. Let $i \in \{1,\dots, |X|-1\}$ be such that $A_i^{\rho} = A_i^{\rho'}$ and $A_{i+1}^{\rho} \neq A_{i+1}^{\rho'}$. Then for all $j \leq i$, no supported path may be in-branching at $A_j$ for either $\rho$ or $\rho'$.
\end{claim}

\begin{proof}
Suppose not. Then, without loss of generality, there exists some supported path $\rho''$ such that $\rho''$ and $\rho$ are in-branching and there exists $j \leq i$ such that $A_j^{\rho} = A_j^{\rho''}$ and $A_{j-1}^{\rho} \neq A_{j-1}^{\rho''}$. Construct supported path $\rho'''$ as follows.
$$\rho'''=(A_0^{\rho''},\dots,A_j^{\rho''},A_{j+1}^{\rho},\dots,A_i^{\rho},A_{i+1}^{\rho'},\dots,A_{|X|}^{\rho'})$$
By construction, $\rho$ and $\rho'''$ are supported paths which are branching. This is a contradiction. Thus our claim is proven.
\end{proof}

\begin{claim}
Let $\rho$ and $\rho'$ be supported in-branching paths. Let $i \in \{1,\dots, |X|-1\}$ be such that $A_i^{\rho} = A_i^{\rho'}$ and $A_{i-1}^{\rho} \neq A_{i-1}^{\rho'}$. Then for all $j \geq i$, no supported path may be out-branching at $A_j$ for either $\rho$ or $\rho'$.
\end{claim}

\begin{proof}
The logic is identical to the proof of Claim 1.
\end{proof}

Together, Claim 1 and Claim 2 state that for every supported path $\rho$ there exists some $i \in \{1,\dots, |X|-1\}$, such that $A_i$ is in $\rho$, with all supported out-branching paths doing so at or above $A_i$ and with all supported in-branching paths doing so strictly below $A_i$. This means that the edge associated with $q(A_i \setminus A_{i+1},A_i)$ belongs to no supported path other than $\rho$. We know from \citet{falmagne1978representation} that any rationalizing $\nu$ must put probability weight on the set of orders ranking $A_i \setminus A_{i+1}$ exactly at the top of $A_i$ equal to $q(A_i \setminus A_{i+1},A_i)$. Since $\rho$ is the unique supported path that contains $q(A_i \setminus A_{i+1},A_i)$, it must be the case that the order $\pi$ associated with $\rho$ must have $\nu(\pi)= q(A_i \setminus A_{i+1},A_i) $. This can be said about all such orders. Thus no pair of supported paths being branching implies that the rationalizing representation must be unique. Thus we have proven our theorem.

\subsection{Proof of Theorem 2}
We begin by proving necessity of the conditions on $\nu$. We proceed by contraposition. Let $\nu$ put strictly positive probability weight on two orders $\pi$ and $\pi'$ satisfying condition 2 of the theorem. This means that there exist $x,y\in X$ such that $x \neq y$ and $U_{\pi}(x)= U_{\pi'}(y)$. Let $A^1$ denote a node in the path corresponding to $\pi$. Similarly, let $A^2$ denote a node in the path corresponding to $\pi'$. Since $x \neq y$ and $U_{\pi}(x)= U_{\pi'}(y)$, there exists some $k < l$ such that $A^1_k \neq A^2_k$ (since $y \not \in U_{\pi'}(x)$ and $x \not \in U_{\pi}(y)$) and $A^1_l = A^2_l$. This means we can find $k< i \leq l$ such that $A^1_i=A^2_i$ and $A^1_{i-1} \neq A^2_{i-1}$. By $U_{\pi}(z) \neq U_{\pi'}(z)$, it must be that $A^1_z \neq A^2_z$. This implies there is some node after $A_i$ at which $\rho$ and $\rho'$ out-branch. Call the first node that does this $A_j$. Thus $A_{i-1}^{1} \neq A_{i-1}^{2}$, $A_{j+1}^{1} \neq A_{j+1}^{2}$, and for all $m \in \{i, \dots, j\}$, $A_m^{1} = A_m^{2}$. This means that $\rho$ and $\rho'$ are a pair of supported branching paths. Thus $\nu$ is not unique by Theorem 1, and by contraposition the conditions on $\nu$ are necessary.

We now show the sufficiency of the conditions on $\nu$. We proceed by contraposition. Suppose that $\nu$ is not unique. Then, by Theorem 1, there is a pair of supported branching paths on the probability flow diagram of $(X,P)$. Recall the definition of branching path. With this definition in mind, we call the length of a branching section $j -i$. Let $l$ be the minimum length of all branching sections of all pairs of supported branching paths. Note that $l$ is well defined because $X$ is finite. Choose a pair of supported paths $\rho$ and $\rho'$ such that $\rho$ and $\rho'$ have a branching section of length $l$. Let $\{A_i, \dots, A_j\}$ be that branching section. Because $l$ is the minimal length of supported branching sections, there is no $k \in \{i+1,j-1\}$ such that $\{A_i, \dots, A_k\}$ or $\{A_k, \dots, A_j\}$ are supported branching sections. We know from \citet{fiorini2004short} and \citet{falmagne1978representation} that the probability flow diagram satisfies inflow equals outflow. Since there is no supported out-branching path in $\{A_i, \dots, A_{j-1}\}$, it must be the case that inflow into $A_i$ equals outflow from $A_j$. 

Let $M_{x,A}$ be the set of linear orders on $X$ that rank $x$ exactly at the top of $A$.
$$M_{x,A}=\{\pi|\pi(X \setminus A) > \pi(x) > \pi(A \setminus x) \}$$
We know from \citet{falmagne1978representation} that $q(x,A) = \nu(M_{x,a})$ for all rationalizing $\nu$. Since $\rho$ and $\rho'$ are supported, the total outflow from $A_j$ is strictly greater than the inflow into $A_i$ from the edge belonging to $\rho$. This means that $\nu$ cannot assign weight onto orders in $M_{A^{\rho}_{i-1} \setminus A_i, A^{\rho}_{i-1} } $ equal to the total outflow from $A_j$.

\begin{claim}
There are two orders, $\pi$ and $\pi'$, satisfying the following.
\begin{enumerate}
    \item $\pi \not \in M_{A^{\rho}_{i-1} \setminus A_i, A^{\rho}_{i-1} }$
    \item $\pi' \in M_{A^{\rho}_{i-1} \setminus A_i, A^{\rho}_{i-1} }$
    \item $\max(\pi,A_j) \neq \max(\pi',A_j)$
    \item $\nu(\pi)>0$ and $\nu(\pi')>0$
\end{enumerate}
\end{claim}
\begin{proof}
Since $\{A_i,\dots,A_j\}$ is a branching section of two supported paths, there are at least two orders whose paths pass through $\{A_i,\dots,A_j\}$ which have positive weight under $\nu$. Further, there must be at least two orders whose paths in-branch at $A_i$ and have positive weight under $\nu$. Similarly, there must be at least two orders whose paths out-branch at $A_j$ and have positive weight under $\nu$. There are two cases.
\begin{enumerate}
    \item There is some supported edge leaving $A_j$ such that no order in $M_{A^{\rho}_{i-1} \setminus A_i, A^{\rho}_{i-1} }$ with positive weight under $\nu$ has a path containing that edge. \\
    Since inflow at $A_i$ equals outflow form $A_j$, it must be the case that some order not in $M_{A^{\rho}_{i-1} \setminus A_i, A^{\rho}_{i-1} }$ has positive weight and has path containing the prior mentioned edge. Call this order $\pi$. Then any $\pi' \in M_{A^{\rho}_{i-1} \setminus A_i, A^{\rho}_{i-1} }$ with $\nu(\pi')>0$ and $\pi$ satisfy the above conditions.
    \item For every supported edge leaving $A_j$, there is some order in $M_{A^{\rho}_{i-1} \setminus A_i, A^{\rho}_{i-1} }$ with positive weight under $\nu$ whose path contains that edge. \\
    In this case, choose some order $\pi \not \in M_{A^{\rho}_{i-1} \setminus A_i, A^{\rho}_{i-1} }$ such that $\nu(\pi)>0$ and the path corresponding to $\pi$ passes through $\{A_i,\dots,A_j\}$. The existence of such an order is guaranteed by inflow equals outflow. By inflow at $A_i$ equals outflow at $A_j$, the path corresponding to $\pi$ passes through $A_j$. Choose some $\pi' \in M_{A^{\rho}_{i-1} \setminus A_i, A^{\rho}_{i-1} }$ such that $\nu(\pi')>0$ and the path corresponding to $\pi'$ does not have the same edge leaving $A_j$ as the path corresponding to $\pi$. The existence of such a $\pi'$ is guaranteed by the supposition. Then $\pi$ and $\pi'$ satisfy the conditions of the claim.
\end{enumerate}
\end{proof}
By the definition of $\rho_{\pi}$ and $\rho_{\pi'}$, both of these paths are branching and supported. Now let $x=\min(\pi,X \setminus A_i)$ and $y = \min(\pi', X \setminus A_i)$. By $\pi \not \in M_{A^{\rho}_{i-1} \setminus A_i, A^{\rho}_{i-1} }$ and $\pi' \in M_{A^{\rho}_{i-1} \setminus A_i, A^{\rho}_{i-1} }$, $x \neq y$. By $A_i \in \rho_{\pi}$ and $A_i \in \rho_{\pi'}$, $U_{\pi}(x) = U_{\pi'}(y)$. Let $z = \max(\pi,A_j)$. By $\max(\pi,A_j) \neq \max(\pi',A_j)$, $U_{\pi}(z) \neq U_{\pi'}(z)$. By $j \geq i$, $\pi(\{x,y\}) > \pi(z)$ and $\pi'(\{x,y\}) > \pi'(z)$. Thus the conditions on $\nu$ hold, and, by contraposition, the sufficiency of the conditions are shown. Thus the theorem is proven.

\subsection{Proof of Theorem 3}
If there are two rationalizing distributions with different supports, then $(X,P)$ is obviously not uniquely rationalizable. All that is left is to show the other direction. Suppose that $(X,P)$ is not uniquely rationalizable. Then $|X| \geq 4$ and the probability flow diagram of $(X,P)$ has supported branching paths. The two algorithms used in the proof of Theorem 1 find two rationalizing distributions with different supports and can be used here to do so. Thus, by contraposition, each rationalizing distribution having the same support implies there is a unique rationalizing distribution.

\bibliographystyle{ecta}
\bibliography{uniqueRUM}

\end{document}